# Modeling mandatory and discretionary lane changes using dynamic interaction networks

Yue Zhang[1], Yajie Zou[1*], Yuanchang Xie[2], Lei Chen[3]

[1]*Key Laboratory of Road and Traffic Engineering of Ministry of Education, Tongji University, Shanghai 201804, China*

[2] *Dept. of Civil and Environmental Engineering University of Massachusetts Lowell One University Avenue, Lowell, MA 01854, USA*

[3] *Department of Mobility and systems, Research Institutes of Sweden, Box 857, 501 15 Borås, Sweden*

# Abstract

A quantitative understanding of dynamic lane-changing (LC) interaction patterns is indispensable for improving the decision-making of autonomous vehicles, especially in mixed traffic with human-driven vehicles. This paper develops a novel framework combining the hidden Markov model and graph structure to identify the difference in dynamic interaction networks between mandatory lane changes (MLC) and discretionary lane changes (DLC). A hidden Markov model is developed to decompose LC interactions into homogenous segments and reveal the temporal properties of these segments. Then, conditional mutual information is used to quantify the interaction intensity, and the graph structure is used to characterize the connectivity between vehicles. Finally, the critical vehicle in each dynamic interaction network is identified. Based on the LC events extracted from the INTERACTION dataset, the proposed analytical framework is applied to modeling MLC and DLC under congested traffic with levels of service E and F. The results show that there are multiple heterogeneous dynamic interaction network structures in an LC process. A comparison of MLC and DLC demonstrates that MLC are more complex, while DLC are more random. The complexity of MLC is attributed to the intense interaction and frequent transition of the interaction network structure, while the random DLC demonstrate no obvious evolution rules and dominant vehicles in interaction networks. The findings in this study are useful for understanding the connectivity structure between vehicles in LC interactions, and for designing appropriate and well-directed driving decision-making models for autonomous vehicles and advanced driver-assistance systems.

# Introduction

Lane-changing (LC) maneuver is one of the most challenging driving tasks for both autonomous and human driven vehicles, since it involves vehicle movements in both longitudinal and lateral directions, which are further constrained by moving vehicles in adjacent lanes. In addition, the motions of the subject vehicle also affect the behavior of surrounding vehicles, creating a dynamically changing, mutually affected, and complex decision-making environment. Therefore, it is necessary to research the interactions among vehicles during lane changes and reveal the underlying interaction patterns and mechanism. Such findings are critical for developing autonomous vehicle (AV) control algorithms that can make reasonable and well-directed LC decisions.

LC interactions have attracted significant attention recently. Zhu et al. (2018) developed a lane-change warning method for autonomous vehicle, considering a safe distance between the subject vehicle and its lead vehicle in the current lane. Liu et al. (2018) proposed a stabilization control strategy for AV LC maneuver, which can account for the surrounding vehicles and their coupled motions. Huang et al. (2021) developed a dynamic LC trajectory planning algorithm for AVs considering the behavior of surrounding vehicles. Since being able to anticipate upcoming driving scenarios is important for AVs, Fernández-Llorca et al. (2020) introduced a two-stream networks approach for predicting the lane changes of surrounding vehicles. Blenk and Cramer (2021) showed that adjacent vehicles in both the target and current lanes had significant impacts on a driver's LC decision-making. Such impacts have also been investigated in many other studies. Tang et al. (2018) and Tang et al. (2019) demonstrated that consideration of the relative distance and velocity between the subject and lead vehicles may improve the accuracy of predicting lane changes. In the LC behavior analysis by Dou et al. (2016), the authors considered the speed differences and positions of the lead and lag vehicles in the target lane with respect to the subject vehicle. Hu et al. (2018) and Ding et al. (2019) used three closest vehicles in the current and two adjacent lanes for predicting lane changes. Li et al. (2021) proposed a survival analysis model to explore the impacts of traffic context such as acceleration, relative speed, and relative distance of four surrounding vehicles on LC duration. It was found that the relative distance between vehicles had a significant impact on LC duration. Although most studies focused on the impacts of surrounding vehicles on the LC vehicle's behavior, Zheng et al. (2013) investigated the effects of an LC vehicle on the lag vehicle in the target lane. To account for the mutual interaction effects among vehicles rather than the one-way effects, Hidas (2005) and Zhang et al. (1998) categorized lane changes into three classes: free, forced, and cooperative lane changes (Wang et al. 2019). Similarly, Ben-Akiva et al. (2006) classified freeway LC operations into three categories: normal, forced, and courtesy lane changes.

LC events have been traditionally divided into mandatory lane changes (MLC) and discretionary lane changes (DLC) (Yang and Koutsopoulos 1996, Pan et al. 2016). MLC occur when a driver must change lane because of a strategic route choice (Van Beinum et al. 2018), such as merging and diverging. DLC are executed when a driver seeks better driving conditions, such as a faster speed, greater following distance, and farther sight distance. Most previous studies considered these two types of lane changes separately (Kesting et al. 2007, Ali et al. 2019, Ali et al. 2020, Li et al. 2021, Wang et al. 2021). Few studies compared and analyzed the differences between these two types of lane changes. Kusuma et al. (2015) found that MLC has smaller acceptable gaps than DLC, and other researchers discovered that the new follower of an LC vehicle in the target lane is more willing to accept small headways (Daamen et al. 2010, Duret et al. 2011). Vechione et al. (2018) conducted statistical analysis to compare the two types of lane changes in terms of several decision variables and only found significant differences in the gap between the subject and preceding vehicles in the original lane. However, this gap variable was an insignificant input to the MLC decision-making model. In another study, MLC were found to be more aggressive and dangerous than DLC (Hao et al. 2020). However, all these studies did not explicitly account for the dynamically changing decision-making environment during the LC process, and more thorough analysis and comparison of MLC and DLC is still needed.

The above studies assumed that vehicle interactions are static since the multi-vehicle interactions are dynamic in nature due to the uncertainty in driver behaviors and the mutual impacts among drivers. Studies have been conducted to account for the dynamics of vehicle interactions. Ji and Levinson (2020) adopted a game theory approach to model real-time vehicle LC interactions. Sun and Kondyli (2010) modeled vehicle LC interactions by adopting the "hand-shaking negotiation" idea in the TCP/IP protocol. These studies analyzed vehicle LC behaviors at every moment, which were computationally intensive and difficult to explain how vehicle interaction patterns evolve during the LC process. To gain insights into vehicle LC interaction patterns, decomposing the LC process into small segments was shown to be a useful approach (Zhang et al. 2021). Higgs and Abbas (2014) developed a two-step algorithm for segmenting and clustering car-following behaviors. After that, the Hidden Markov Model (HMM) was adopted as a common segmentation method to uncover dynamic internal states in continuous driving behavior sequences (Wang et al. 2018, Zhang and Wang 2019, Zhang et al. 2021). However, two important issues remain unsolved: What are the interactive relationships among vehicles in a lane-changing process? How to quantitatively model such interactions?

Vehicle interactions refer to the correlations among vehicle behaviors during driving. Statistics such as the Pearson's correlation coefficient (PCC) and Spearman's correlation coefficient (SCC) have been widely used to quantify the linear correlations between variables. However, they cannot identify direct associations due to the fact that they rely only on information of co-occurring events (Zhao et al. 2016). To solve this problem, partial correlations (Ryali et al. 2016) were used to evaluate direct associations by eliminating the effects of additional variables. However, these correlation measures could only detect linear associations and may ignore important nonlinear relationships that likely exist given the complexity of vehicle interactions. In order to consider nonlinear associations, mutual information (MI) (Kraskov et al. 2004) based on information theory was proposed to evaluate the dependence between two random variables. Unlike the PCC, SCC, and partial correlation, MI is not limited to real-valued random variables. It is more general and depends on how similar the variable distributions are. Like PCC and SCC, MI has the overestimation problem and cannot assess direct associations due to the only information of joint probability (Zhao et al. 2016). To accurately consider both nonlinear and direct associations, conditional mutual information (CMI) was proposed and it has been a powerful tool widely used across areas such as genetic engineering (Zhang et al. 2012), natural language processing (Adaimi and Thomaz 2019), and data annotation (Saad and Mathiak 2013). CMI is also suitable for evaluating the association between two vehicles. However, an LC process usually involves multiple mutually-affected vehicles simultaneously, and only measuring two-vehicle associations is not enough to model the interactions among vehicles involved. Furthermore, to understand multiple vehicle interactions, it is necessary to quantify the interaction intensity of each pair of vehicles in the LC process, identify interaction patterns, and investigate how such patterns evolve. To account for multi-vehicle interaction quantitatively, this study characterizes interactions by innovatively constructing graph theory-based networks (Sebastian et al. 2009) and further proposes a framework to quantitatively analyze the dynamic interaction networks for MLC and DLC, respectively.

To summarize, this study has two main objectives. The first one is to model the dynamic vehicle interactions during lane changes. Given the multiple vehicles surrounding a subject vehicle in an LC process, a nonlinear measure is proposed to quantify the interaction intensity between each pair of vehicles. Based on the measured interaction intensities, LC interaction networks are identified and constructed together with explorations of the network evolution during the LC process. The second one is to compare the characteristics of MLC and DLC based on the proposed framework. Two types of lane changes are compared from the perspectives of interaction intensity and the dynamic transition of interaction networks. The fundamental differences between MLC and DLC and their working mechanism have been revealed.

## Methodology

The framework proposed in this paper is shown in Figure 1. Firstly, LC interaction scenarios are extracted from naturalistic driving data. Then homogenous states are segmented by the Gaussian HMM. Given these states, the vehicle interactions are measured by CMI. Finally, a graph theory-based method is utilized for structure representation to describe the interactions between individual pairs of vehicles simultaneously.

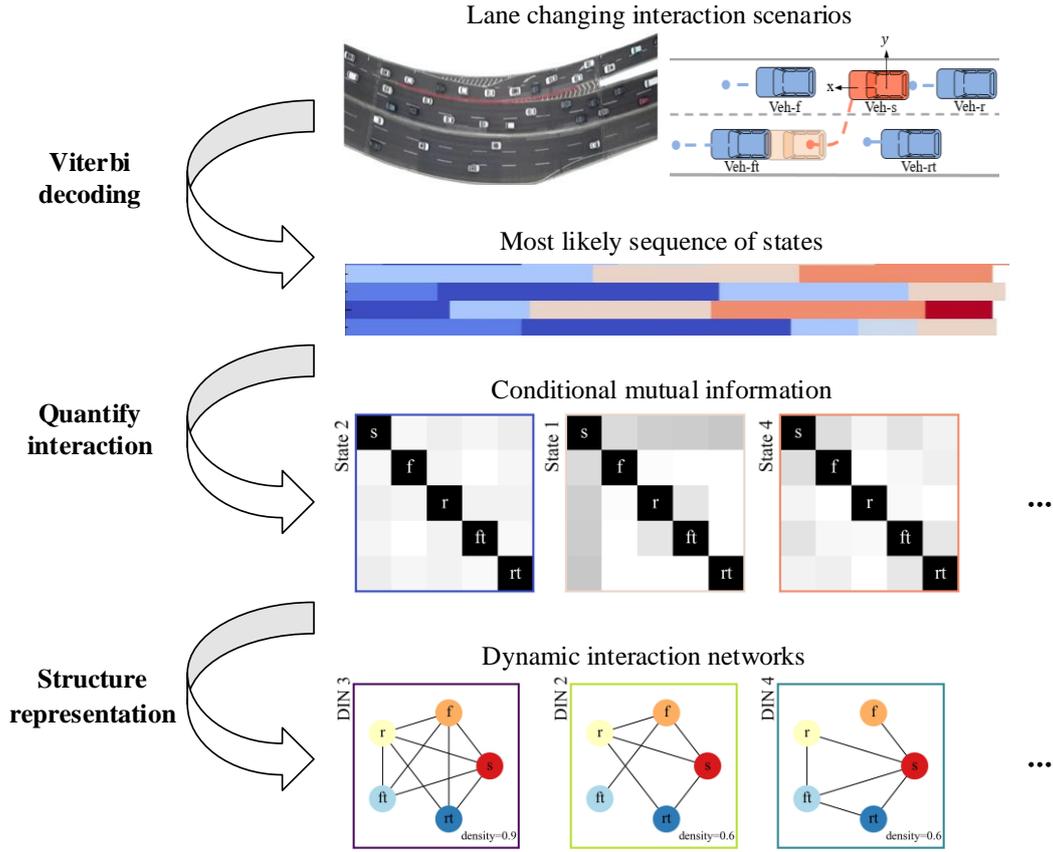

Figure 1  Flowchart of the proposed graph theory-based framework

## Gaussian Hidden Markov Model (HMM)

Dynamic driving behavior can be modeled as a Markov process since it only depends on the most recent observation (Zhang and Wang 2019). As a flexible and efficient model, the HMM decomposes sequences into homogeneous and discrete hidden states. These hidden states often have important physical meaning in applications to facilitate model interpretation (Rabiner 1989). In this study, the Gaussian HMM is adopted to analyze the dynamics of LC vehicle interactions which is illustrated in Figure 2.

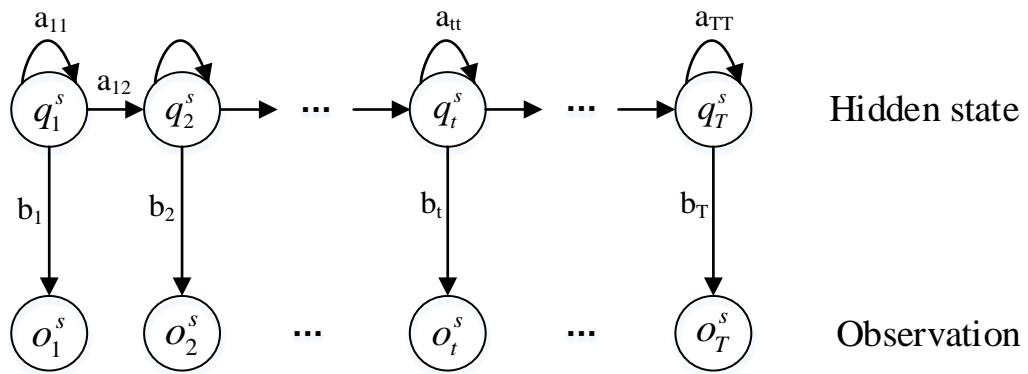

Figure 2 The schematic diagram of the proposed Gaussian HMM

Let $O = \{\{o_t^s\}_{t=1}^T\}_{s=1}^S$ be the time series describing LC interaction scenarios, where $T$ is the LC duration and $S$ is the number of LC events. $o_t^s = \{x_m, y_m\}_{m=1}^M$ is a $2 \times M$ dimensional observation at time $t$ for the $m$-th vehicle in lane-changing event $s$, where $M$ is the number of vehicles of interest (or nodes of the dynamic interaction network) and $x, y$ are the longitudinal and lateral positions of the $m$-th vehicle respectively. For each observation in the time series, there is only one hidden state corresponding to it. Hidden states are expressed as $Z = \{z_1, z_2, \cdots, z_k\}$, where $K$ is the number of hidden states. Let $Q = \{\{q_t^s\}_{t=1}^T\}_{s=1}^S$ be a 1$^{st}$ order Markov chain, where $q_t^s \in Z$ is the discrete hidden state at time $t$ in lane-changing event $s$. The state transition matrix $A = \{a_{ij}\}$ represents the transition probability from state $i$ to state j. And the transition probability for (i, j) pairs can be expressed as Eq. (1).

$$a_{ij} = P(q_{t+1} = Z_j | q_t = Z_i), \text{ s.t. } i \geq 1, j \leq k \text{ and } \sum_{j=1}^{k} a_{ij} = 1 \ \forall \ i \qquad (1)$$

The probability of an observation $o_t^s$ generated from state $q_t^s$ is called emission probability and represented by $b(t)$. Since the observations in this study are continuous, we assume the emission probability to be a multivariate Normal distribution as expressed by Eq. (2),

$$b(t) = P(o_t^s | q_t^s = z_k) = \mathbb{N}(\mu_{z_k}, \Sigma_{z_k}), \quad 1 \leq k \leq K \quad 1 \leq t \leq T \qquad (2)$$

where $\mu_{z_k}, \Sigma_{z_k}$ are parameters mean and covariance, respectively.

All the parameters in this model are estimated by the Expectation-Maximization algorithm. To decompose the observations into homogeneous and discrete states, we utilize the Viterbi algorithm to find the corresponding optimal hidden state sequence $Q^*$ given the model. Then, some hidden state properties are calculated to describe dynamic network characteristics for a state $z_k$, which are defined as follows.

**Frequency** is the number of transitions from other states (excluding the state $z_k$) to the state $z_k$.

**Occupancy proportion** is the proportion of time spent on the state $z_k$ in total lane-changing duration, which is defined as Eq.(3).

$$Occupancy\ proportion(z_k) = \frac{\sum_{t=1}^{ST_s} I(q^*(t) = z_k)}{ST_s} * 100 \qquad (3)$$

Where $I(q^*(t) = z_k)$ is a Kronecker delta, which equals 1 if the current state $q^*(t)$ is $z_k$. $ST_s$ is the total time duration of the $s$ LC events.

**Mean lifetime rate** is the ratio of the average time of the state $z_k$ continuous maintenance to the average duration of all LC events.

## Quantify interaction intensity

In order to better uncover the associations among vehicles, CMI is proposed to quantify nonlinear and direct relationships. Identifying dependencies or associations between vehicles is the premise of constructing interaction networks.

### *Conditional mutual information (CMI)*

In information theory, mutual information (MI) is a measure of the mutual dependence between two random variables. Let $X$ and $Y$ be two random variables, MI represents the shared information when one of X and Y is obtained. For continuous random variables X and Y, MI is defined in Eq. (4),

$$I(X;Y) = \int_Y \int_X p(x,y) \log\left(\frac{p(x,y)}{p(x)p(y)}\right) dx\, dy \qquad (4)$$
$$= H(X,Y) - H(X|Y) - H(Y|X)$$

where $p(x,y)$ denotes the joint probability distribution of X and Y, $p(x)$ and $p(y)$ are the marginal probability distributions of X and Y, respectively, and H is the Shannon entropy. A limitation of MI is that it ignores the impacts of other random variables and cannot measure the direct relationship, which may lead to overestimation (Zhang et al. 2012). Therefore, CMI is utilized to measure direct dependence by considering conditional probability. CMI for continuous variables $X, Y, Z$ is described in Eq. (5),

$$I(X;Y|Z) = \int_Z \int_Y \int_X p(x,y,z) \log\left(\frac{p(x,y|z)}{p(x|z)p(y|z)}\right) dx\,dy\,dz \qquad (5)$$
$$= H(X|Z) + H(Y|Z) - H(X,Y|Z)$$

where X and Y are scalar variables, and Z is a vector of multi-dimensional variables. In this study, X represents the longitudinal or lateral position of one vehicle, Y represents the longitudinal or lateral position of another vehicle, and Z includes the longitudinal and lateral positions of all vehicles except the two covered by X and Y. Such an input arrangement is to ensure that the dependence between X and Y is not affected by Z, especially when both X and Y are weakly associated with Z.

CMI is used to evaluate the direct association between pairs of vehicles in the following dynamic interaction networks. Note that CMI is non-negative. A higher CMI value implies a stronger dependency between X and Y. CMI is estimated using the nearest-neighbor entropy estimator defined in Eq. (6),

$$\hat{I}_{XY|Z} = \psi(k) + \frac{1}{n}\sum_{i=1}^{n}\left[\psi(k_i^z) - \psi(k_i^{xz}) - \psi(k_i^{yz})\right] \qquad (6)$$

$$\psi(x) = \frac{d}{dx}\ln\Gamma(x) \qquad (7)$$

where $n$ is the sample length, $k$ is the number of nearest neighbors in the joint space of $x \otimes y \otimes z$ with the local length scale $\epsilon_i$, $k_i^{xz}$ is the number of points with a distance strictly smaller than $\epsilon_i$ in the subspace $x \otimes y$, and $k_i^{yz}$ and $k_i^z$ are computed in the same way.

*Conditional independence test*

This section is to test the conditional independence hypothesis for CMI. The null and alternative hypotheses are described in Eq. (8) and Eq. (9), respectively. According to Eq. (5), it is clear that $I(X;Y|Z) = 0$ only if $X \perp\!\!\!\perp Y \mid Z$.

$$H_0: X \perp\!\!\!\perp Y \mid Z \qquad (8)$$
$$H_1: X \not\!\perp\!\!\!\perp Y \mid Z \qquad (9)$$

To test the conditional independence of X and Y, a nearest-neighbor permutation test (Runge 2018) is used in this paper. Traditionally, the CMI surrogates used to simulate independence are generated by randomly permuting $x$ in $\{x_i, y_i, z_i\}_{i=1}^n$, where $n$ is the number of samples (Doran et al. 2014). However, this approach gets rid of all dependencies between $x$ and $y$ as well as between $x$ and $z$, resulting in $X \perp\!\!\!\perp Y \mid Z$ not being tested accurately. Hence, a local permutation test based on the nearest-neighbor search is applied to preserve the dependence between $x$ and $z$. This approach estimates the null distribution by applying the CMI estimator based on the surrogates generated by the local permutation test. When the estimated value of the CMI surrogates is greater than or equal to the CMI of the original data, the $p$ value can be derived using Eq. (10) and Eq. (11),

$$\hat{I}_b = \hat{I}(x^*; y|z) \qquad (10)$$

$$p = \frac{1}{B}\sum_{b=1}^{B} \mathbf{1}[\hat{I}_b \geq \hat{I}(x; y|z)] \qquad (11)$$

where $x^*$ is drawn from the nearest-neighbor permutation, the detailed process of generating the samples can be found in (Runge 2018). $B$ is the number of generated CMI surrogates and $\mathbf{1}$ is an indicator function. Based on the $p$ values obtained from Eq. (11), we can determine whether there exists an interaction or not. Note that $p$ values only represent the significance of interactions. On the other hand, CMI values describe the closeness of interactions between vehicles.

## Interaction network construction

We propose a graph theory-based approach to explicitly model the statistical dependencies and interactions between vehicles during lane changes. The interaction network is defined as an undirected graph $\mathcal{G} = (\mathcal{V}, \mathcal{E}, \mathcal{w})$, where $\mathcal{V} = \{v_1, v_2, \cdots v_M\}$ is a set of nodes, $\mathcal{E} \in \{\langle v_a, v_b\rangle : v_a, v_b \in \mathcal{V}, v_a \neq v_b\}$ is a set of edges, and $\mathcal{w} = \{w_{e_{ab}}, e_{ab} \in \mathcal{E}, a \neq b\}$ is the set of weights for those edges. In this study, a node represents a vehicle in the LC scenario, an edge $e_{ab} = \langle a, b\rangle$ with weight $w_{e_{ab}}$ represents the interaction and its intensity between vehicles $a$ and $b$, and the network $\mathcal{G}$ represents the entire LC scenario. For each pair of vehicles $(v_a, v_b)$, we need to check whether there is a significant association between vehicles $a$ and $b$ according to the $p$ value obtained from Eq. (11). If $p \leq 0.05$, an edge $e_{ab}$ with weight $w_{e_{ab}}$ will be created and added to $\mathcal{E}$. The weight $w_{e_{ab}}$ is CMI, which is calculated using Eq. (5). If $p > 0.05$, there is no association between vehicles $a$ and $b$. The statistics commonly used in network analysis include network density and node degree. Network density describes the closeness between nodes in network $\mathcal{G}$. It is the ratio of the existing number of edges to the number of total possible edges in a network as defined in Eq. (12),

$$D = \frac{2|\mathcal{E}|}{|\mathcal{V}|(|\mathcal{V}| - 1)} \qquad (12)$$

where $|\mathcal{V}|$ is the number of nodes, and $|\mathcal{E}|$ is the number of edges in the graph. The weighted node degree measures the importance of individual nodes. Assuming that $A_a = \{e_{a1}, e_{a2} \cdots e_{aM}\}$ is a set of edges adjacent to node , the weighted node degree $deg(v_a)$ is defined in Eq. (13).

$$deg(v_a) = \sum_{b \in \mathcal{V}} w_{e_{ab}} e_{ab} \;, e_{ab} \in A_a \qquad (13)$$

# Data description

This section introduces the LC data and the data acquisition together with pre-processing methods used in this paper.

## Data collection

The INTERACTION (Zhan et al. 2019) dataset is used in this study, which contains vehicle trajectories collected by Unmanned Aerial Vehicles (UAVs) from multiple countries such as the U.S.,

Germany, and Bulgaria. Compared with other open datasets, the INTERACTION dataset contains more LC scenarios with close interactions between vehicles in congested traffic. This study selects an on-ramp section in China for analysis as shown in Figure 3. The chosen video is 94.62 minutes long and contains 10,359 vehicles. The data refresh frequency is 10 Hz. To ensure consistency, both MLC and DLC are considered based on the westbound traffic (i.e., vehicles above the concrete median in Figure 3).

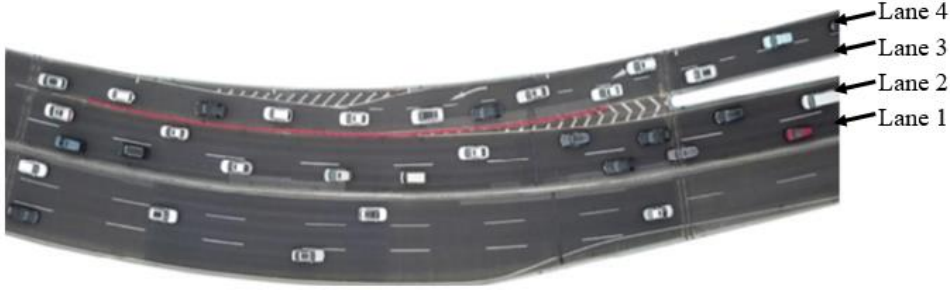

Figure 3 The lane changing interaction scenario at a freeway on-ramp in the INTERACTION dataset

## Lane-changing interaction scenario extraction

To exclude the impacts of external factors such as road geometry when comparing MLC and DLC, this study extracts MLC and DLC at the same location. The schematic diagrams of MLC and DLC are shown in Figure 4. Note that the MLC in this study are only for merging.

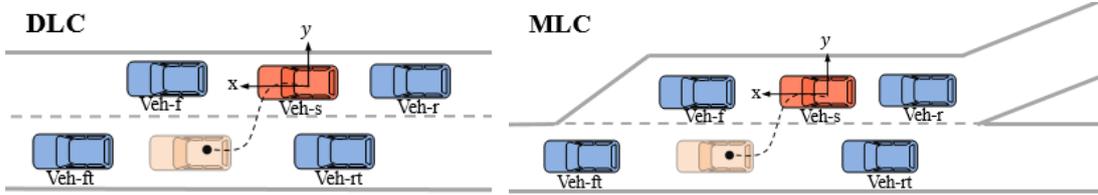

Figure 4 Schematic diagrams of interaction scenarios generated by two kinds of LC behavior

*Identify LC events*

The LC events are extracted by the following steps.

    Step 1: Model lane boundaries with polynomial functions. Since the selected road segment is on a horizontal curve, a quadratic polynomial function is established to fit each lane boundary. The fitted function is a curve along the lane markers separating two lanes, which is important for identifying LC events. Given the $x$ value of a point on the curve, its $y$ coordinate can be expressed as $y = f(x)$.

    Step 2: Identify LC vehicles and their paths. For MLC, when a vehicle crosses lane marker $l$ and its lane ID changes from Lane 4 to Lane 3 (see definition in Figure 3) at time $t_c$, the vehicle is identified as an LC vehicle *veh-s*. For DLC, when a vehicle crosses lane marker $l$ and its lane ID changes from Lane 2 to Lane 1 or Lane 1 to Lane 2 at time $t_c$, the vehicle is identified as an LC vehicle *veh-s*.

    Step 3: Find LC start and end times. The start ($t_s$) and end ($t_e$) times of an LC process are determined based on vehicle trajectory records (Zhao et al. 2017) using Eqs. (14)-(15),

$$t_s \leftarrow \min(\{t|Dis(t) - 1 > 0\}), t \in [\,t_s, t_c\,] \quad (14)$$
$$t_e \leftarrow \min(\{t'|Dis(t') + 1 > 0\}), t' \in [\,t_c, t_e\,] \quad (15)$$

where $Dis(t) = |y_t - f_l(x_t)|$ is the lateral distance from *veh-s* to lane marker *l*. The duration of an LC process is $t_e$-$t_s$. The duration distributions of the extracted MLC and DLC events are shown in Figure 5. Note that larger MLC durations are observed because LC vehicles in this particular dataset tend to stay very close to the left lane markers for a long time before crossing one.

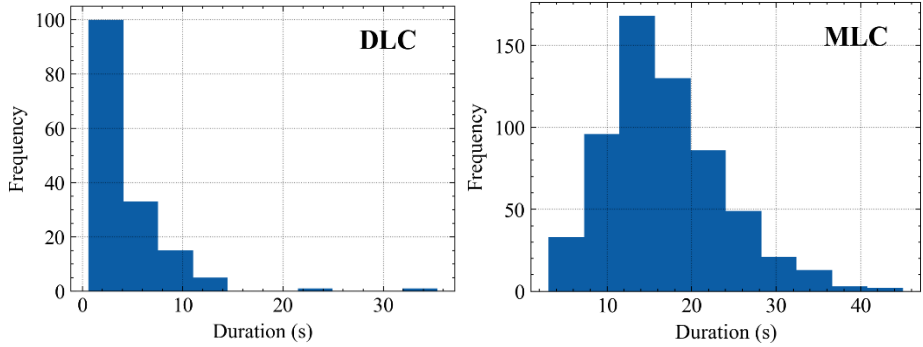

Figure 5 Duration distribution of all extracted LC events

*Match surrounding vehicles*

As shown in Figure 4, the trajectories of all related surrounding vehicles and the subject vehicle are taken as the inputs. The related surrounding vehicles include the lead and lag vehicles in the original lane and the target lane. Therefore, the number of interest vehicles $M = 5$ and $o_t^s$ in Eq. (2) is a 2 × 5 vector. In the absence of any of the lead and lag vehicles, the corresponding values in $o_t^s$ are set to some constants, which equal the coordinates of the lower-left corner of the video frame. This point is in the opposite direction of the traffic under investigation and has no effects on the extracted LC scenarios.

Traffic flow can affect LC location, intensity, and accepted gaps. This paper focuses on LC behavior under congested traffic, specifically levels of service E and F. According to the Highway Capacity Manual (HCM), LC events are extracted when the traffic density is greater than 35 (pc/mi/ln) (i.e., 56.35 (pc/km/ln)).

# Results and discussion

## Interaction State Learning and Temporal Analysis

The Gaussian HMM is used to decompose LC scenarios into homogenous interaction states. To determine the optimal $K$ value in Eq. (2), the log-likelihood value is used to evaluate the performance of the model by changing the value of $K$ from 2 to 20 with an increment of 1. Figure 6 shows that the optimal $K$ values are 10 for DLC and 7 for MLC.

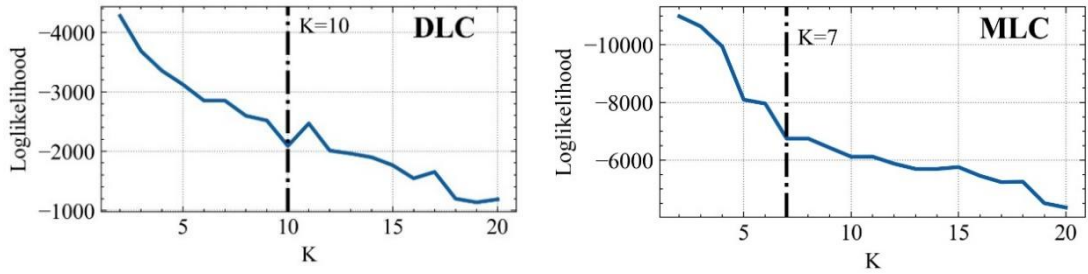

Figure 6 Selection process of optimal state number K

Figure 7 shows the dynamic interaction states automatically discovered by the Gaussian HMM. The left and right columns show the segmentation results for DLC and MLC, respectively. In each column, different colors are used to represent different states. Note that, the colors in the two columns are independent (i.e., the same color in the left and right columns does not mean the same state). The states are sorted in descending order based on their occupancy proportions and are named State 1, State 2, and so on.

Figure 7a) and 7b) show the decomposing results of LC events and how an LC vehicle transits from one homogenous interaction state to another. Taking the first DLC event in Fig. 7a) as an example, this DLC event is decomposed into three interaction states, and the corresponding vehicle transitioned from interaction state 8 to 6 and ended in state 5. For the first MLC event in Fig. 7b), the MLC event is decomposed into six interaction states. Based on the order of occurrence, these six states are 5, 2, 3, 6, 1, and 4. Figure 7c) and 7d) consist of multiple subplots and each subplot shows the CMI matrix for an interaction state. In each CMI matrix, each grid/cell represents the CMI value of two interactive vehicles except for those diagonal cells. For example, the cell of the first row and the second column is the CMI value of the vehicle pair (*veh-s*, *veh-f*). The darker the cell color, the greater the CMI value (i.e., more intense interactions between the two vehicles). Clearly, the MLC interactions are closer than DLC interactions. The average CMI for MLC is 0.26, compared to 0.13 for DLC. Figure 7g) and 7h) show that the average lifetime rates are about 0.3 (<<1) for DLC and 0.25 (<<1) for MLC, indicating that the temporal persistence of a state is much shorter than the length of an LC process. It also means that an LC event can be divided into several homogeneous states (on average $1/0.3 \approx 3$ states for DLC and $1/0.25 = 4$ states for MLC).

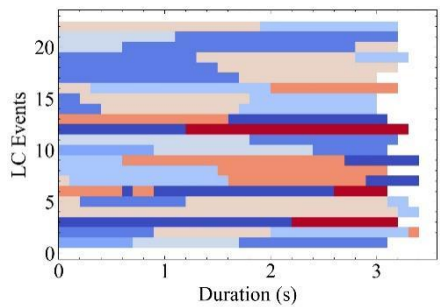
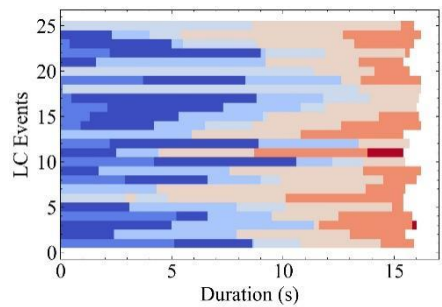

a) Time evolution of dynamic states in some DLC

b) Time evolution of dynamic states in some MLC

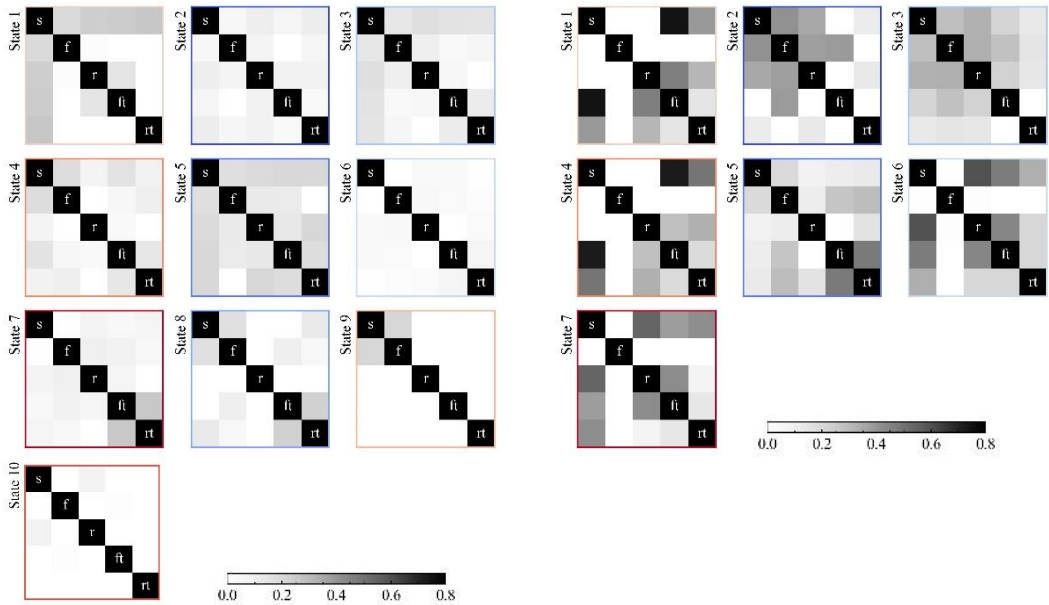

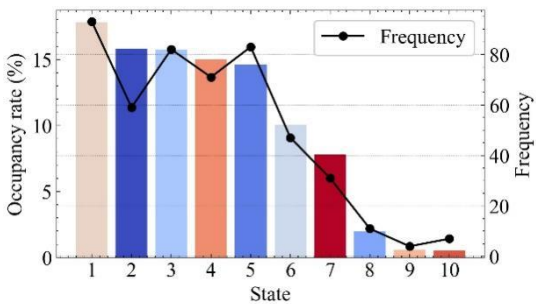

c) Estimated CMI of all DLC [1]

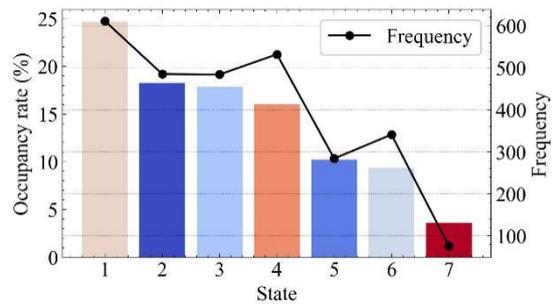

d) Estimated CMI of all MLC [1]

e) Occupancy proportion and frequency of each state in DLC

f) Occupancy proportion and frequency of each state in MLC

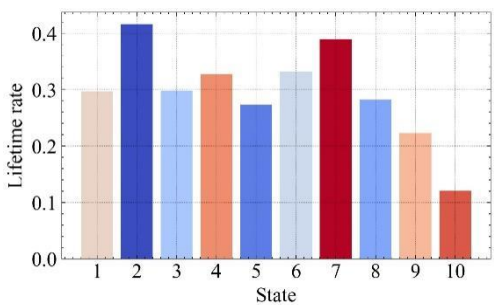

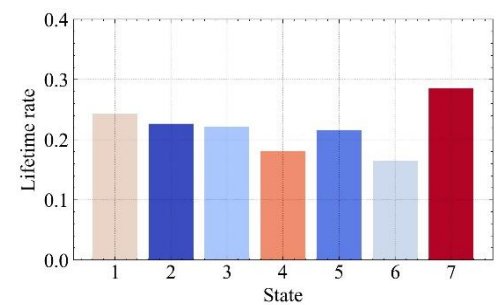

g) Lifetime rate of each state in DLC

h) Lifetime rate of each state in MLC

[1] There is no corresponding value for the diagonal in c) and d), because this study focuses on the association between vehicles rather than the autocorrelation of vehicles.

Figure 7 Dynamic states identified by Gaussian HMM in DLC and MLC

## Dynamic interaction network (DIN) analysis

To further characterize the connectivity patterns associated with the identified interaction states (see

Figure 7), a graph theory-based approach is applied to generate DIN for each pattern. We will also describe the salient features of the dynamic interaction networks (DIN) in DLC and MLC. To better illustrate how various DIN structures can be used to describe lane changes, some states with a similar network structure but different edge weights are combined into a same DIN. According to Figure 7 e), since the top eight states constitute 98% of the total occupancy rate and the probabilities of transferring to the remaining two states are almost zero, states 9 and 10 are eliminated. Five DINs are identified from each type of LC events (i.e., DLC and MLC), as shown in Figure 8 c) and d). In these two figures, the thick edge connecting the two nodes indicates that the interaction between the corresponding two vehicles is intense. These DINs are further explained below to gain insight into the DIN structure and interaction patterns between vehicles. According to network density (see Eq. (12)), these DINs are classified into two groups: dense network (density>0.7) and sparse network (density ≤ 0.7), where the threshold value of 0.7 is the mean network density of all DINs. The following part presents in more detail the results for DLC and MLC, respectively, together with a comparative discussion.

**DLC:** Both DIN 2 and DIN 4 are dense networks, where all vehicles are connected except for one pair of vehicles. In DIN 2, although the number of connections is high, the intensity of the connection is very weak (see states 3 and 7 in Figure 7 c)). In this case, *veh-r* and *veh-rt* are disconnected from each other, corresponding to an LC scenario where both *veh-r* and *veh-rt* are stably driving in their respective lanes and do not affect each other. Compared to DIN 2, the interactions of DIN 4 are stronger (see state 5 in Figure 7 c)). The structure of DIN 4 shows that *veh-f* and *veh-rt* are disconnected. This is because the gap between these two vehicles is large.

DIN 1, DIN 3, and DIN 5 are sparse networks. Among them, DIN 1 (states 2, 4, and 6 in Figure 7 c)) has the weakest interactions on all edges. This corresponds to *veh-s* with a strong LC intention, where it has large gaps to the lead and lag vehicles in the original lane, and only needs to take care of the lead and lag gaps in the target lane (the gaps concerning *veh-ft* and *veh-rt*). The behavior of *veh-r* is very similar to that of *veh-s* (*veh-s* and *veh-r* are only connected to *veh-ft* and *veh-rt*), where the vehicle is affected by the low speed of *veh-f* and may want to change lane. DIN 3 (state 1 in Fig. 7c)) represents LC scenarios where the interactions between *veh-s* and other vehicles dominate. In those scenarios, the interactions between *veh-s* and surrounding vehicles are stronger than the interactions among surrounding vehicles. DIN 5 (state 8 in Fig. 7c)) shows an obvious community structure, in which *veh-r* is completely isolated while the community of the remaining four vehicles has good connectivity. This represents an LC scenario where *veh-r* is far away from other vehicles.

In DINs 1~4 of DLC, *veh-r* is always associated with *veh-ft* and *veh-f*. This is because of the slow speed of *veh-f*, *veh-r* not only pays attention to *veh-f* in the current lane, but also monitors *veh-ft* in the target lane to evaluate the potential benefits of changing lane (Keyvan-Ekbatani et al. 2016).

**MLC:** Different from DLC, *veh-s*, *veh-f* and *veh-r* in the original lane (i.e., the acceleration lane) in MLC scenarios all have imminent LC needs. There are three dense networks in the MLC case, including DIN 1, DIN 3, and DIN 5. In all the three DINs, *veh-ft* is not fully connected with other vehicles, suggesting that more critical constraints in MLC come from *veh-f*, *veh-s*, *veh-r*, and *veh-rt*. In DIN 1, *veh-s* has strong interactions with all the surrounding vehicles, while *veh-f* has very weak interactions with *veh-r* and *veh-ft* (states 1 and 4 in Figure 7 d)). This scenario usually occurs when *veh-s* is crossing the lane marker and *veh-r* is also about to change lane (thus paying attention to *veh-rt*). In this case, there is no *veh-f* in the acceleration lane (since *veh-s* is the first vehicle in the acceleration lane) and the interaction between *veh-f* and other vehicles is almost zero. In DIN 3 (state 3), *veh-ft* and *veh-rt* are disconnected. This is because when vehicle-s is inserted, *veh-rt* needs to evaluate the utility of *veh-s* insertion and follow *veh-s*, thus turning its attention from *veh-ft* to *veh-s* and resulting in no association between *veh-ft* and *veh-rt*. DIN 5 (state 5) often appears in the first stage of an MLC process, where *veh-f* and *veh-s* start to change lanes at the same time. Veh-r is still relatively far from the junction entrance (i.e., merging point) and does not need to change lane immediately. Veh-ft and Veh-rt coordinate closely (i.e., strong interactions) to share the gap in the target lane.

DIN 2 (state 2) and DIN 4 are sparse networks. The intense interactions in DIN 2 mostly occur in the original lane, and these interactions are mainly due to car-following behaviors in the original lane. The network structure of DIN 4 in MLC is very similar to that of DIN 3 in DLC, but its interaction is stronger (states 6 and 7 in Figure 7 d)).

In all the DIN structures for MLC, *veh-s* is always connected with *veh-f* and *veh-r*. This is because the lag vehicle needs to pay close attention to the lead vehicle and keep a safe distance. Therefore, the car-following behavior in the acceleration lane has generated intense interactions for MLC. Also, *veh-s* and *veh-rt* are connected in all DINs. This can be explained as the LC vehicle (*veh-s*) needs to pay more attention to *veh-rt* to determine whether *veh-rt* is going to yield or not.

**Comparison:** In both DLC and MLC, *veh-s* and *veh-rt* always maintain an interactive relationship, which is consistent with the results of previous LC studies based on traditional methods. In terms of network connectivity, there are three dense network structures in MLC and only two in DLC. As for interaction intensity, vehicles in MLC interact with each other more closely than in DLC. Therefore, the interaction closeness of MLC scenario is significantly higher than that of DLC.

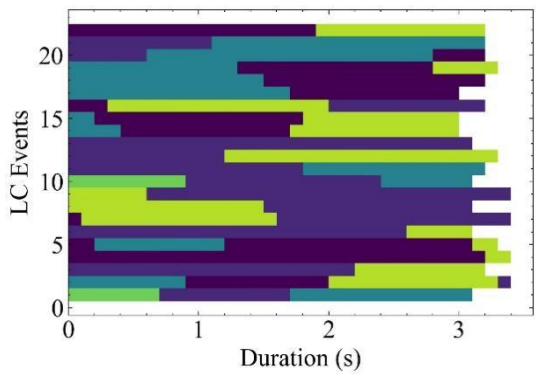
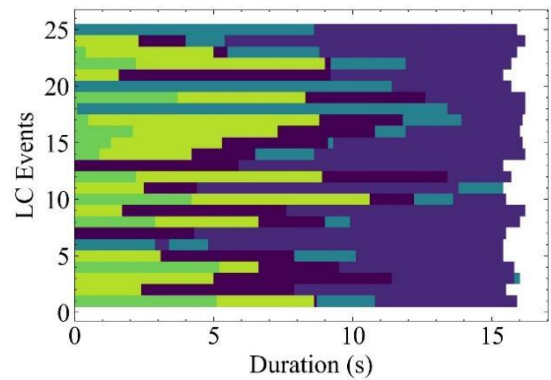

a) Time evolution of DINs in some DLC  b) Time evolution of DINs in some MLC

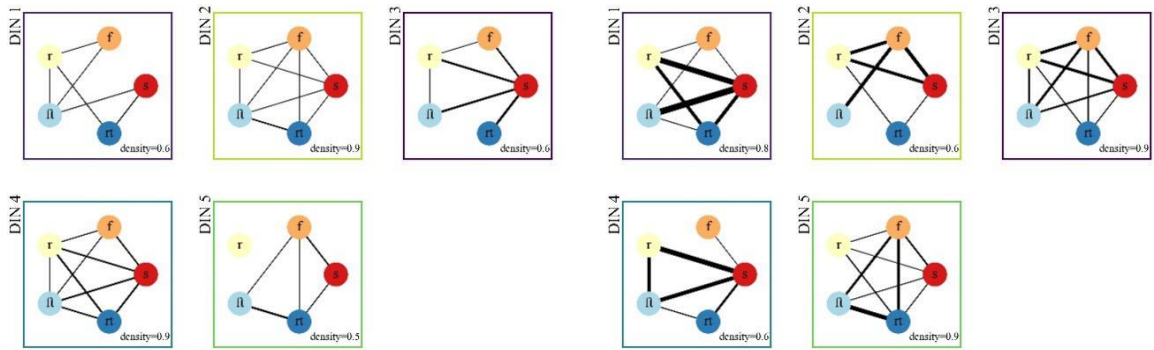

c) DIN structure of DLC        d) DIN structure of MLC

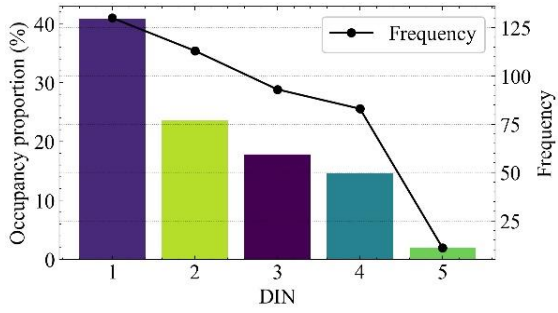   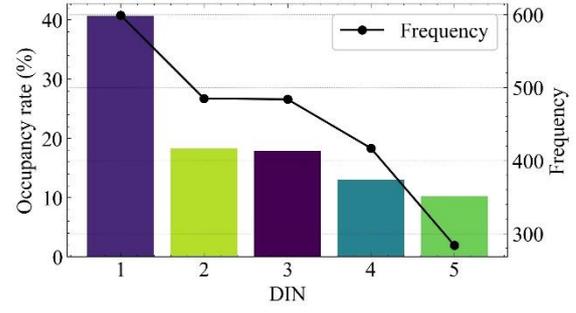

e) Occupancy proportion and frequency of each DIN in DLC        f) Occupancy proportion and frequency of each DIN in MLC

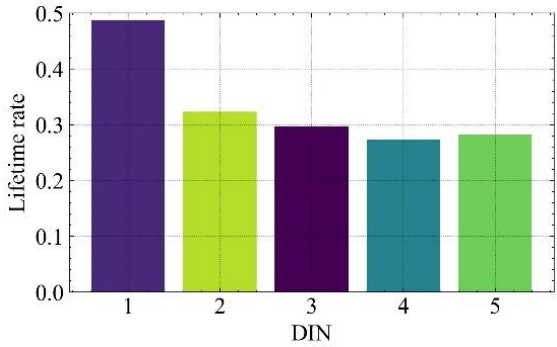   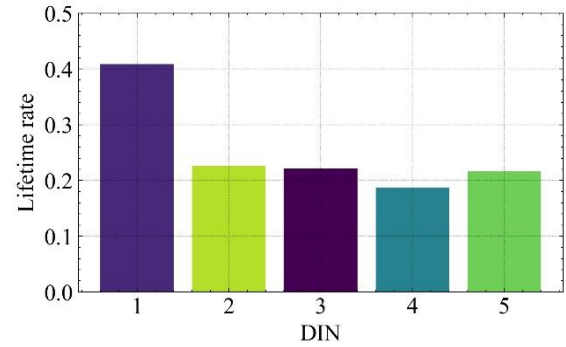

g) Lifetime rate of each DIN in DLC        h) Lifetime rate of each DIN in MLC

Figure 8 Dynamic interaction networks identified in DLC and MLC

## Comparison of dynamic interaction network evolution in MLC and DLC

Based on the results in Figure 8 for MLC and DLC, temporal properties of the DINs were further analyzed and major findings are summarized as follows:

1. In LC scenarios, the connectivity and interaction intensity between vehicles clearly change dynamically. As Figure 8 a) and b) show, LC events move from one DIN to another constantly over time and the time duration of each DIN varies across LC events. Take the first DLC event (Figure 8 a)) as an example, the LC vehicle experienced three DINs in the order DIN 5 – DIN 1 – DIN 4, which shows a trend from weakly to strongly connected networks. When *veh-s* starts the lane change, *veh-r* is far away from the community of other vehicles and acts independently (i.e., DIN 5). As *veh-s* approaches the lane marker, *veh-r* gets closer and gradually joins the LC interactions, where *veh-r* starts to pay attention to the locations of vehicles in the target lane and shows behavior similar to those of *veh-s* (DIN 1). Towards the end of the LC process, *veh-s* interacts closely with other surrounding vehicles when crossing the lane marker and entering the target lane. Meanwhile, *veh-r* is also approaching and crossing the lane marker and exhibits interaction characteristics very similar to those of *veh-*

*s* (DIN 4). Different from previous studies which assumed static interaction relationships during lane changes, this study clearly shows that the associations between vehicles change over time.

2. The comparative study of Figure 8 a) and 8 b) suggests that the sequence of DINs seems less randomly in MLC than that in DLC. Figure 8 a) shows that in DLC, the positions of each DIN in different DLC events are random, where a DIN may appear at the beginning or the end of DLC. While for MLC, they typically consist of DINs in the following order: DIN 5- DIN 2- DIN 3- DIN 4- DIN 1 (Figure 8 b)). Note that some MLC events do not include all five types of DINs. The differences here between MLC and DLC are interesting but not very surprising, since DLC can happen under a wide range of conditions, while MLC scenarios are mostly similar to each other.

3. Another major difference is that MLC experience more frequent DIN changes than DLC. In other words, MLC on average experience more DINs than DLC. According to Figure 8 e) and 8 f), DINs are more evenly distributed among different types for MLC than for DLC. In particular, DIN 5 makes up less than 5% of total DIN occurrences in DLC, while in MLC the minimum occupancy proportion is 10%. Figure 8 g) and 8 h) show that DIN 1 (which has the highest occupancy proportion) in DLC has an average lifetime of 0.5 LC events, while DIN 1 (also has the highest occupancy proportion) in MLC can only survive 0.4 LC events. Besides, the lifetime rates of other DINs in MLC are all shorter than those in DLC, indicating that DINs in MLC last shorter durations than those in DLC. The average lifetime rate of all DINs in DLC is 0.34, while the corresponding average in MLC is 0.24. This means that each DLC on average will experience three DINs (1/0.34) and each MLC will experience about four DINs (1/0.24).

## Identify critical vehicles

Based on the identified DINs, the importance of each node (i.e., vehicle) can be quantified using the weighted node degree (see Eq. (13)), which describes a node's influences on other vehicles' behavior during lane changes. A larger weighted node degree means that the corresponding vehicle is more critical.
Table 1 summarizes the weighted node degree values. In each row (i.e., DIN), the value that is significantly higher than the rest is highlighted in bold font. Overall, *veh-s* is the most critical vehicle (i.e., dominates) in nearly all DINs, because the LC maneuvers of *veh-s* lead to traffic perturbations and increase the interactions among all vehicles. In MLC, *veh-f* also dominates two common DINs: DIN 2 (1.18) and DIN 3 (1.06). Since *veh-f* leads *veh-s*, when *veh-s* intents to change lane in an MLC case, so does *veh-f*. Therefore, vehicles in the target lane need to also pay attention to *veh-f*, which is closer to the entrance of the ramp. In addition, *veh-f* affects the behavior of vehicles behind it in the queue in the acceleration lane (including *veh-s* and *veh-r*). In contrast, there are no obvious dominant vehicles in DINs 1 and 2 in DLC, which most likely is due to the discretionary nature of DLC, as *veh-s* has more flexibility in deciding when and where to change lane. Such flexibility is also reflected by the randomness in DIN changes during DLC (see discussion in the previous section, bullet item #2). In addition, Table 1 shows that the weighted node degrees (which are calculated by interaction intensities) among vehicles in MLC are significantly higher than those in DLC, which again supports the findings regarding interaction intensity in section *Dynamic interaction network*

*analysis*.

Table 1 Weighted degree of each vehicle in each DIN

|  | DIN | s | f | r | ft | rt |
|---|---|---|---|---|---|---|
| **DLC** | 1 | 0.24 | 0.12 | 0.25 | 0.25 | 0.24 |
|  | 2 | 0.29 | 0.38 | 0.22 | 0.39 | 0.32 |
|  | 3 | **0.94** | 0.22 | 0.41 | 0.39 | 0.26 |
|  | 4 | **0.77** | 0.42 | 0.66 | 0.63 | 0.59 |
|  | 5 | 0.30 | 0.32 | 0.00 | 0.33 | **0.39** |
| **MLC** | 1 | **1.74** | 0.00 | 1.00 | 0.82 | 0.87 |
|  | 2 | 0.89 | **1.18** | 0.86 | 0.39 | 0.25 |
|  | 3 | 0.97 | **1.06** | 1.02 | 0.72 | 0.41 |
|  | 4 | **1.29** | 0.01 | 0.93 | 0.95 | 0.39 |
|  | 5 | 0.52 | 0.88 | 0.35 | 0.86 | **1.06** |

# Conclusions

This paper develops a novel framework combining HMM and graph structure to better understand the fundamental mechanism of MLC and DLC and the key differences between them. The Gaussian HMM reveals heterogeneous states in an LC process. The dynamic interaction networks further quantify the complex interactions among vehicles involved and the temporal properties of such interactions based on real-world LC data. The key findings and contributions of this paper are summarized below:

(1) It introduces a CMI measure to quantitatively describe the nonlinear and direct associations between vehicles. The dynamic interaction networks in an LC process are then identified based on CMI, and semantic interpretation is performed based on real-world LC data.
(2) Different from previous studies that assumed static interaction relationships in LC, this study clearly demonstrates the dynamically changing nature of interactions among vehicles during LC processes using Gaussian HMM. It further classifies such networks into different groups and reveals the evolving patterns of these dynamic interaction networks.
(3) MLC and DLC are dissected and compared from the perspective of dynamic interaction networks. It is found that MLC involve more complicated maneuvers and more intense interactions than DLC, as manifested by the intense interaction and frequent transitions of network structures. More specifically for interaction intensity, MLC not only have stronger interactive vehicle pairs (average CMI is 0.26 for MLC vs. 0.13 for DLC), but also have a larger number of intense network structures (3 for MLC vs. 2 for DLC). For network structure transition, the dynamic interaction network structure on average changes twice in a DLC process and three times in an MLC process.
(4) The transitions from one dynamic interaction network structure to another during DLC seem random and are difficult to predict, while some clear patterns for MLC are shown. This probably is because the subject vehicle (*veh-s*) in DLC has much more flexibility to choose when and where to change lanes than in MLC.
(5) The proposed framework provides a novel network-based approach to quantitatively measure

and segment the dynamically changing interactions among vehicles and identify the most critical vehicle in LC scenarios.

# Future work

With the development of the network-based approach and quantitative methods, as well as the investigation of lane changes based on realistic data, this study opens doors to many interesting directions.

One of the most interesting directions is autonomous vehicles. Autonomous vehicles are equipped with sophisticated sensors that measure the surrounding traffic, generating a huge amount of driving behavior data to be analyzed. The proposed framework and method can be used to model such data and better understand the interactions between autonomous vehicles and the surrounding traffic under a variety of driving conditions. The findings can help researchers develop safer and more robust autonomous vehicle control algorithms. For example, the comparison results of MLC and DLC in this study can help autonomous vehicles understand the intent of surrounding human-driven vehicles and develop optimal and customized control strategies. Moreover, the method of identifying critical vehicles facilitates to solve the high-dimensional challenges in the process of generating autonomous driving test scenarios (Feng et al. 2021).

Another interesting topic for applying the proposed method is to analyze highly interactive driving scenarios, such as multi-vehicle encounters at intersections, LC behavior in weaving areas and before work zones, and pedestrian-vehicle conflicts on urban roads. Scenarios could include interactions among human-driven vehicles, as well as mixed traffic with both autonomous and human-driven vehicles. New data sets such as Waymo Open Dataset (Ettinger et al. 2021) could be used to verify and enhance the proposed method.

Furthermore, the proposed method for identifying critical vehicles can help to formulate efficient communication protocols and optimize wireless channel ussage by reducing redundant transmission (Sebastian et al. 2009). For example, the vehicle may receive only the context information of critical vehicles in the scenario.